\documentclass[prb,twocolumn]{revtex4}
\usepackage{amsmath}
\usepackage{bm}
\usepackage{epsfig}
\usepackage{amssymb}

\begin{document}
\newcommand{\figwidth}{0.95\columnwidth}
\newcommand{\ffigwidth}{0.4\columnwidth}

\title{Critical parameters for the disorder-induced metal-insulator transition in FCC and BCC lattices}

\author{Andrzej Eilmes$^{1}$, Andrea M. Fischer$^{2}$, Rudolf A.\ R\"{o}mer$^{2}$}
\affiliation{
  $^1$Department of
  Computational Methods in Chemistry,\\ Jagiellonian University,
  Ingardena 3, 30-060 Krak\'{o}w, Poland\\ $^2$Department of Physics
  and Centre for Scientific Computing, University of Warwick, Coventry,
  CV4 7AL, United Kingdom}

\date{\today}

\begin{abstract}
We use a transfer-matrix method to study the disorder-induced metal-insulator transition. We take isotropic nearest- neighbor hopping and an onsite potential with uniformly distributed disorder. Following previous work done on the simple cubic lattice, we perform numerical calculations for the body centered cubic and face centered cubic lattices, which are more common in nature. We obtain the localization length from calculated Lyapunov exponents for different system sizes. This data is analyzed using finite-size scaling to find the critical parameters. We create an energy-disorder phase diagram for both lattice types, noting that it is symmetric about the band center for the  body centered cubic lattice, but not for the face centered cubic lattice. We find a critical exponent of approximately 1.5-1.6 for both lattice types for transitions occurring either at fixed energy or at fixed disorder, agreeing with results previously obtained for other systems belonging to the same orthogonal universality class. We notice an increase in critical disorder with the number of nearest neighbors, which agrees with intuition.
\end{abstract}


\maketitle

\section{Introduction}
\label{sec-intro}
The disorder-induced metal-insulator transition (MIT) and the concept of Anderson localization \cite{And58,EcoC70,EcoC72,Eco72,LicE74} have been studied extensively for more than forty years. The scaling theory of localization \cite{AbrALR79} provides a very successful approach for non-interacting electrons. According to its predictions, a disorder driven MIT occurs in three-dimensional systems, i.e.\ beyond a critical amount of disorder $W_{\rm c}$ all eigenstates localize. For smaller disorder, extended states exist in the system.

For the simple cubic (SC) lattice and uniform disorder distribution, the critical disorder and the critical exponent have been successfully calculated using the transfer-matrix method (TMM).\cite{PicS81a,KraM93} Highly accurate recent studies \cite{Mac94,SleO99a,MilRSU00,OhtSK99} report $W_{\rm c} = 16.54 \pm 0.02$ and the critical exponent $\nu = 1.57 \pm 0.02$. \cite{SleO99a} However, direct diagonalization results based on energy-level statistics \cite{ZhaK95c,ZhaK97,MilR98,MilRS00} and multi-fractal analysis \cite{MilRS97,Mil00} give a smaller $\nu = 1.44 \pm 0.2$. Furthermore, experimental results report yet smaller values of $\nu \gtrsim 1.0$. \cite{StuHLM93,WafPL99,ItoWOH99,ItoHBH96,WatOIH98} Many of these observed discrepancies can be explained by the attainable limits on system sizes, temperature and statistical averages in the above results, nevertheless, the quest for an accurate determination of the critical parameters at the Anderson transition is not yet complete.

The SC lattice is, in addition, not very common \cite{AshM76} in nature. The only element known to adopt it is the alpha phase of polonium; most metals exhibit body centered (BCC) or face centered (FCC) cubic lattices. Although the localization properties for an FCC lattice have been studied recently \cite{LudTED05} for a vibrational problem, to the best of our knowledge no critical parameters have been reported for an electronic Anderson transition in BCC and FCC lattices. In order to fill this gap, we use in the present paper the TMM and finite-size scaling (FSS) to calculate the critical parameters for the MIT in the BCC and FCC lattices. Of course, one should expect no change in the critical exponent, as all systems considered here belong to the same orthogonal universality class. On the other hand, because of the different numbers of nearest neighbors in SC ($Z=6$), BCC ($8$) and FCC ($12$) lattices, the values of the critical disorder $W_{\rm c}$ and energy $E_{\rm c}$ may be different.   Hence our study tests and reconfirms universality while at the same time allowing to see how the non-universal parameters of the transition change with increasing coordination number.

\section{Numerical approach}
\label{sec-numapp}

\subsection{The Transfer-Matrix Approach to\\ the Anderson Model of Localization}
\label{sec-model}
\label{sec-TMM}

To model the MIT in the 3D system we use the standard Anderson Hamiltonian
\begin{equation}
  \bm{\mathrm{H}} = \sum_i \epsilon_i \left| i \right\rangle \left\langle i \right| -
  \sum_{i \neq j} t_{ij} \left| i \right\rangle \left\langle j
  \right|. 
\label{hamilt}
\end{equation}
The orthonormal states $\left| i \right\rangle$ correspond to
electrons located at sites $i = (x,y,z)$ of a cubic lattice with periodic 
boundary conditions. The hopping integrals $t_{ij}$ are non-zero only for
$i,j$ being nearest-neighbors and the energy scale is set by choosing
$t_{ij} = 1$. The disorder in the model is incorporated into the
diagonal energies $\epsilon_i \in [-W/2, W/2]$, randomly distributed according to the uniform distribution with width $W$.


In order to compute the localization length $\lambda$ of the wave function,
we use the TMM for quasi-1D bars of cross section
$M \times M$ and length $L \gg M$.\cite{PicS81a,KraM93,SleO99a,MilRSU00} The Schr\"{o}dinger equation 
$\bm{\mathrm{H}} \psi = E \psi$ for the Hamiltonian given by Eq.\ (\ref{hamilt}) is written in the TMM form:
\begin{multline}
\left(
\begin{array}{c}
\psi _{l+1} \\
\psi _l
\end{array}
\right) 
=\bm{\mathrm{T}}_l\left(
\begin{array}{c}
\psi _l \\
\psi _{l-1}
\end{array}
\right),\\
 =\left(
\begin{array}{cc}
-\bm{\mathrm{C}}_{l+1}^{-1}(E\bm{\mathrm{1}}-\bm{\mathrm{H}}_l) &  
-\bm{\mathrm{C}}_{l+1}^{-1}\bm{\mathrm{C}}_{l} \\
\bm{1} & \bm{0}
\end{array}
\right)
\left(
\begin{array}{c}
\psi _l \\
\psi _{l-1}
\end{array}
\right),
\label{eq-tmm}
\end{multline} 
where $\psi_l, \bm{\mathrm{H}}_l$ and $\bm{\mathrm{T}}_l$ denote the wave function, Hamiltonian matrix and transfer matrix of the $l$th slice of the bar, respectively. 
$\bm{1}$ and $\bm{0}$ denote unit and zero matrices. 
The localization length $\lambda(M,W)=1/\gamma_{\rm min}$ at energy
$E$ is determined by the smallest Lyapunov exponent $\gamma_{\rm min}>0$
obtained as an eigenvalue of the product of transfer matrices
$\tau_L = \bm{\mathrm{T}}_L \bm{\mathrm{T}}_{L-1} \ldots \bm{\mathrm{T}}_2 \bm{\mathrm{T}}_1$, where $L$ is increased until the
desired accuracy is achieved.\cite{Ose68} The reduced localization length
may then be calculated as $\Lambda_M(W)=\lambda(M,W)/M$.

$\bm{\mathrm{C}}_{l}$ and $\bm{\mathrm{C}}_{l+1}$ are the connectivity matrices describing the connections of the $l$th slice to slices $l-1$ and $l+1$.\cite{SchO92} Element $c_{jk}$ of the connectivity matrix  equals $1$ if the site $j$ in one slice is connected to the site $k$ in the other; otherwise $c_{jk}=0$.
In the case of the SC lattice, each site has only one connection to the succeeding (preceding) layer;
therefore all $\bm{\mathrm{C}}_l$ are unit matrices and the 
transfer matrix $\bm{\mathrm{T}}_l$ reduces to the most often used form\cite{minussign}
\begin{equation}
\label{eq-tmmred}
\boldsymbol{\mathrm{T}}_{l} = \left(
\begin{array}{cc}
-(E\bm{1}-\bm{\mathrm{H}}_l) & -\bm{1} \\
\bm{1} & \bm{0}
\end{array}
\right).
\end{equation} 
For BCC and FCC lattices the connectivity matrices take a more complicated form, but with purely diagonal disorder, i.e.\ no disorder in the hopping integrals $t_{ij}$, they are constant  so that the inverse $\bm{\mathrm{C}}_l^{-1}$ needs to be calculated only once at the beginning of the TMM calculations for a given size $M$. Nevertheless, the additional need to multiply all states at each step of the TMM with a dense matrix $\bm{\mathrm{C}}^{-1}$ reduces the speed of the calculation and hence restricts the attainable system sizes.

\subsection{The Lattice Structures}
\label{sec-latt}

\begin{figure}
  \centering
  (a)\includegraphics[width=\ffigwidth]{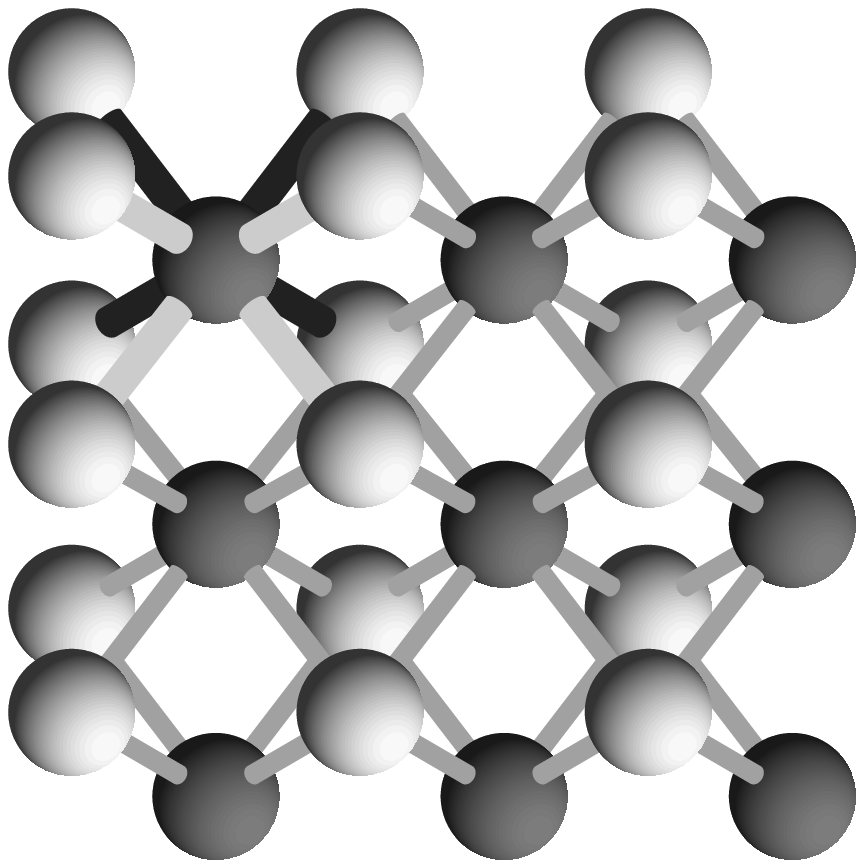}
  (b)\includegraphics[width=\ffigwidth]{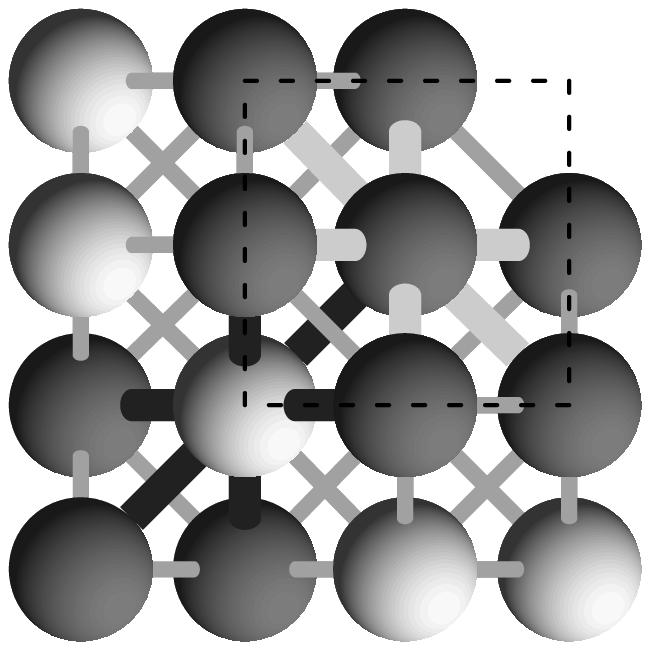}
  \caption{
    (a) 3 layers of the 3D BCC lattice along a $\langle 100\rangle$ lattice vector. Light gray spheres mark the 1st and 3rd layer, dark gray ones indicate the central layer. Lines between layers denote the connections between lattice sites. The connections to the upper-left sphere in the central layer are emphasized by broad lines, illustrating its 8 neighbors. The 4 thick light gray lines connect 1st and the central layer, the black ones go from the central to the 3rd layer.
    (b) Structure of 3 layers of the 3D FCC lattice along a $\langle 111\rangle$ lattice vector. The broken lines mark the cubic unit cell of the lattice. Sites in the 1st and 3rd layer are dark gray; sites in the central layer are light gray. The thick light gray lines represent connections between lattice sites in the {\em same} layer for one particular site. The thick black lines represent connections between lattice sites in {\em  neighboring} layers for another site. The thin lines indicate connections to other sites. Some sites in the upper-right corner are removed for clarity.}
\label{fig-bccfcc3D}
\end{figure}

The structure of the BCC lattice is displayed in Fig.\ \ref{fig-bccfcc3D}(a).
The construction of the TMM quasi-1D bar proceeds along a $\langle 100\rangle$ vector. In this case each site within the slice is connected to four sites
in the preceding slice and to four sites in the succeeding one. There are no
connections between sites within the slice, which means that the Hamiltonian
matrix $\bm{\mathrm{H}}_l$ is a diagonal matrix of energies $\epsilon_i$.
We use periodic boundary conditions in both transversal directions, which results in the connectivity matrix for a slice of $M \times M$ sites being singular for all even $M$, thus restricting the system sizes we can use.  Using a helical boundary condition \cite{ZhaK99} in one or two direction provides the same singularities and hence offers no advantage.

Fig.\ \ref{fig-bccfcc3D}(b) shows the structure of the FCC lattice. It proved convenient to construct the TMM bar along a $\langle 111\rangle$ vector, so the subsequent layers of the bar are close packed. Within the layer each site has six connections to nearest neighbors. In addition there are three connections to the preceding and three connections to the succeeding layer. The resulting connectivity matrix can be inverted for each size of the $M \times M$ TMM slice but only when we use a mix of periodic boundary conditions in one direction and helical boundary conditions in the other. See appendix for examples of the connectivity matrices for system size $M=3$. 

\subsection{Finite-Size Scaling}
\label{sec-fss}

The MIT is characterized by a divergent correlation length, so that at fixed energy $E$, $\xi(W)\propto |W-W_{\rm c}|^{-\nu}$ and at fixed disorder $W$, $\xi(E)\propto |E-E_{\rm c}|^{-\nu}$, where
$\nu$ is the critical exponent and $W_{\rm c}$, $E_{\rm c}$ are the critical disorder and energy, respectively, at which the MIT occurs. \cite{KraM93} In the following discussion we shall assume the case of fixed energy and varying disorder; the converse case of fixed disorder and varying energy proceeds analogously.

In order to extract the critical parameters from the calculated values of
$\Lambda_M(W)$, one applies the FSS procedure outlined in Ref.\ \onlinecite{MacK83}. The correlation length for the infinite system $\xi$ may be obtained from the
localization lengths for finite system sizes $\Lambda_M(W)$ by using
the one-parameter scaling law $\Lambda_M=f(M/\xi)$.\cite{Tho74}
The FSS can be performed numerically by minimizing the deviations of the data
from a common scaling curve. The critical parameters are then obtained by fitting the $\xi$ values as obtained from FSS.
Better numerical accuracy for the FSS procedure can be achieved by fitting directly the raw data from TMM calculations using the method applied previously to the TMM data for the 3D SC lattice. \cite{SleO99a,MilRSU00} We introduce a set of fit functions which include two kinds of corrections to scaling, (i) nonlinearities of the $W$ dependence of the scaling variables and (ii) an irrelevant scaling variable which accounts for a shift of the point at which the $\Lambda_M(W)$ curves cross. We use\cite{MilRSU00}
\begin{equation}
  \label{eq-Slevin}
  \Lambda_M=\tilde{f}(\chi_{\rm r} M^{1/\nu}, \chi_{\rm i} M^{y})
  \quad ,
\end{equation} 
where $\chi_{\rm r}$ and $\chi_{\rm i}$ are the relevant and irrelevant scaling variables respectively. The function $\Lambda_M(W)$ is then Taylor expanded
\begin{eqnarray}
  \label{eq-Slevin2}
  \Lambda_M&= &\sum_{n=0}^{n_{\rm i}} \chi_{\rm i}^n M^{n
    y}\tilde{f}_n(\chi_{\rm r} M^{1/\nu}) \quad ,\\
%
  \tilde{f}_n&= &\sum_{k=0}^{n_{\rm r}} a_{nk} \chi_{\rm r}^k M^{k/\nu}
  \quad .
\end{eqnarray}
Nonlinearities are taken into account by expanding $\chi_{\rm r}$ and
$\chi_{\rm i}$ in terms of $w=(W_{\rm c}-W)/W_{\rm c}$ up to order $m_{\rm
  r}$ and $m_{\rm i}$, respectively,
\begin{equation}
  \label{eq-Slevin-Var}
  \chi_{\rm r}(w)=\sum_{m=1}^{m_{\rm r}} b_m w^m , \quad \chi_{\rm
    i}(w)=\sum_{m=0}^{m_{\rm i}} c_m w^m \quad ,
\end{equation}
with $b_1=c_0=1$. The expansions in the fit functions are carried out up to orders 
$n_{\rm i}, n_{\rm r}, m_{\rm r}, m_{\rm i}$ which are adjusted to the specific
data and should be kept as low as possible, while giving the best fit to the data and FSS plot and minimizing the errors for critical parameters $W_{\rm c}$ and $\nu$. The Levenberg-Marquardt method  was used to perform the non-linear fit.\cite{MilRSU00,PreFTV92}. 

We emphasize that this FSS procedure assures the divergence of $\xi$ and hence it is not the divergence itself but rather the quality of how the model fits the computed reduced localization lengths $\Lambda_M$ which determines the validity of the scaling hypothesis.

\section{Calculations and results}
\label{sec-calc}

\subsection{Phase Diagrams}
\label{sec-phasediag}

\begin{figure}
  \centering
  \includegraphics[width=\figwidth]{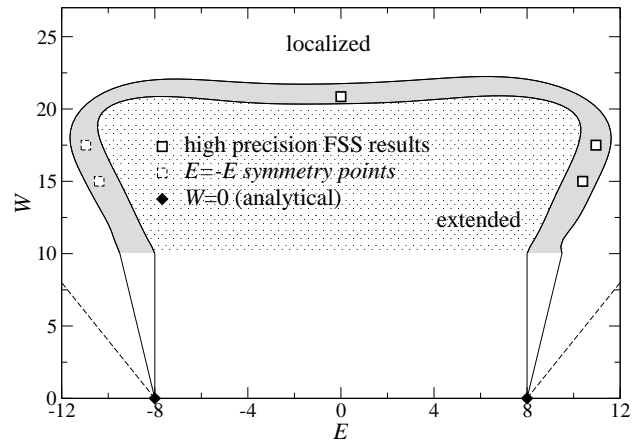}
  \caption{
    Phase diagram for the BCC lattice. The dark gray region represents the approximate location of the phase boundary. Its edges (the solid black lines) were 
    determined by comparing localization lengths with errors $\le 10 \%$ for system sizes $M=7$ and $M=9$ in the $(E,W$) plane. The solid squares $(\Box)$ are points calculated by performing high-precision FSS on localization data with an error $\le 0.1 \%$. The dashed squares are reflections of the solid squares in the $E=0$ axis. The diamonds $(\blacklozenge)$ denote the band edges at $W=0$. They have been joined to the phase boundary edges calculated for higher disorders as a guide to the eye. The dashed lines are the theoretical band edges $\pm (Z + W/2)$, where $Z$ is the coordination number. The light gray, shaded area in the center contains extended states; states outside the phase boundary are localized. Error bars are within symbol size for $(\Box)$. }

\label{fig-bccphasediag}
\end{figure}

\begin{figure}
  \centering
  \includegraphics[width=\figwidth]{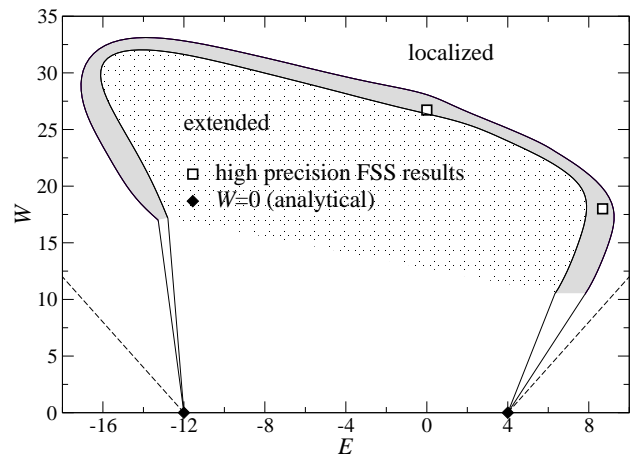}
  \caption{
    Phase diagram for the FCC lattice. Symbols, lines and shaded areas have the same meaning as in Fig.\ \ref{fig-bccphasediag} with the diamonds $(\blacklozenge)$ representing the band edges $-12$ and $4$ at zero disorder \cite{AshM76}. Error bars are within symbol size. }
    
\label{fig-fccphasediag}
\end{figure}

Fig.\ \ref{fig-bccphasediag} 
and Fig.\ \ref{fig-fccphasediag} 
show the phase diagrams for the BCC and FCC lattices, respectively. Originally a grid of $W$ versus $E$ values was created with separation $\Delta E$, $\Delta W =0.5$. At each point the nature of the electronic wavefunction was determined by comparing the reduced  localization lengths $\Lambda_M$ calculated for system sizes $M=7$ and $M=9$ with error $\le 10 \%$. If $\Lambda_9 > \Lambda_7$ ($<$) at the same values of $E$ and $W$ then we identify the point $(E,W)$ in the phase diagram as extended (localized). The edges of the phase boundary were obtained by averaging separately over the three extended and localized points $(E,W)$ nearest to the boundary and then connecting such averages using a spline fit. We do not obtain data points for lower disorder values, as the fluctuations in the Lyapunov exponents, due to the small system sizes, become too big; higher values of disorder smooth out these fluctuations. 

A striking difference between the phase diagrams is that for the BCC lattice the phase boundary is symmetric about the line $E=0$, whereas for the FCC lattice it is not. This is due to the bipartiteness of the BCC lattice which consists of two SC sublattices, one displaced half the distance along a body diagonal of the other. Hence for any site in one sublattice, its nearest neighbors are in the other sublattice. Such connections result in states coupled by a bipartite symmetry transformation --- which is exact for the case of no diagonal disorder --- with eigenenergies of the same magnitude but opposite sign having approximately the same localization lengths; this produces a symmetric phase diagram. The FCC lattice is non-bipartite, so such a symmetry in its phase diagram is not observed. 

\subsection{Critical Parameters at $E=0$}
\label{sec-critdis}

The TMM calculations were performed for system sizes up to $M = 15$. In order to examine the localization properties at the band center for the BCC lattice and the barycenter\cite{barycenter} for the FCC lattice, we set $E=0$ in Eq.\ (\ref{eq-tmm}). A value of the critical disorder $W_{\rm c}$ was approximated using the phase diagrams described above and then localization lengths $\lambda$ were calculated for a range of $W$ close to this approximate value with accuracy ranging from $0.1 \%$ for small system sizes $M$ to about $0.14 \%$ for the largest. Let us remark that we use the term critical disorder to indicate that there are no further extended states at $E=0$ for disorders $W>W_{\rm c}$; extended states may still exist for $W>W_{\rm c}$ at other energies $E$, as shown in Fig.\ \ref{fig-fccphasediag}.

\begin{figure}
  \centering
  \includegraphics[width=\figwidth]{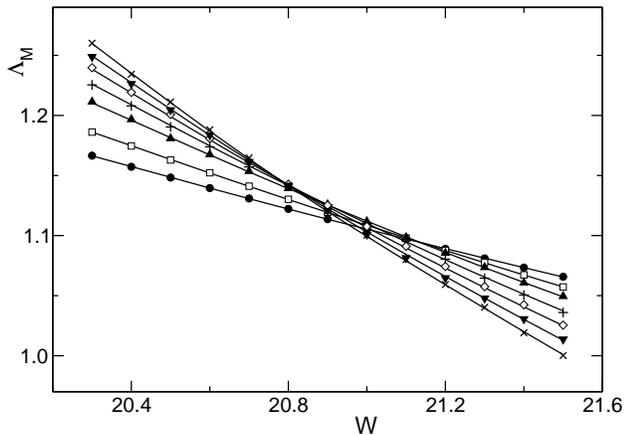}
  \caption{
    Reduced localization lengths $\Lambda_M$ versus disorder $W$ for BCC lattice. 
    System sizes $M$ are $3 (\bullet), 5 (\Box), \ldots, 15 (\times)$.
    Error bars are within symbol size.
    Lines are fits to the data given by Eqs.\ (\ref{eq-Slevin}) -- (\ref{eq-Slevin-Var})
    with $n_{\rm r}=3, n_{\rm i}=2, m_{\rm r}=3, m_{\rm i}=1$. }
\label{fig-bcc-data}
\end{figure}

\begin{figure}
  \centering
  \includegraphics[width=\figwidth]{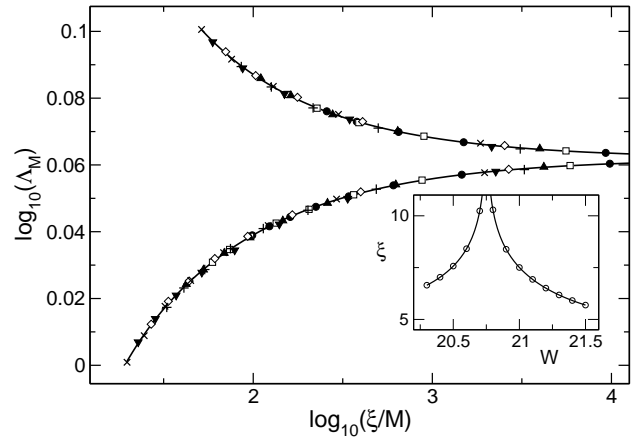}
  \caption{
    Scaling function (solid line) and scaled data points for the BCC lattice and
    $n_{\rm r}=3, n_{\rm i}=2, m_{\rm r}=3, m_{\rm i}=1$. Symbols denote the same values of $M$ as in Fig.\ \ref{fig-bcc-data}. Inset: Dependence of the scaling parameter $\xi$ on the
    disorder strength $W$ for the $13$ $W$ values shown in Fig.\ \ref{fig-bcc-data}. In all cases error bars are within symbol size. }
\label{fig-bcc-scal}
\end{figure}

The reduced localization lengths for the BCC lattice are displayed in Fig.\ \ref{fig-bcc-data}. 
Note how the crossing point of the curves shifts with changing $M$. In most cases this indicates the need for an irrelevant scaling variable introduced via non zero values of $n_{\rm i}$ and $m_{\rm i}$ in Eqs.\ (\ref{eq-Slevin2}) and  (\ref{eq-Slevin-Var}). Fig.\ \ref{fig-bcc-scal} 
shows the results of the scaling procedure for $n_{\rm r}=3, n_{\rm i}=2, m_{\rm r}=3, m_{\rm i}=1$. The scaling curve exhibits localized and extended branches as expected for the MIT. Divergence of the scaling
parameter $\xi$ at $W \approx 20.75$ indicates the critical value of the disorder.  
Table \ref{tab:bcc}(a)
\begin{table}
  \centering
\caption{Critical parameters for the MIT in the BCC lattice. All errors quoted are standard errors.
(a) 3 examples of FSS results with varying $n_{\rm r}, n_{\rm i}, m_{\rm r}, m_{\rm i}$ at fixed energy $E=0$. We use $91$ data points, equally spaced in the indicated intervals (cp.\ Fig.\ \ref{fig-bcc-data}), for each set of $n_{\rm r}, n_{\rm i}, m_{\rm r}, m_{\rm i}$. Varying $n_{\rm r}, n_{\rm i}, m_{\rm r}, m_{\rm i}$, we obtain $77$ best fit models in order to produce the indicated averages. 
(b) Similar FSS results obtained for $3$ out of $19$ best fit models from $82$ non-equally spaced data points at fixed $W=15$ for the indicated energy intervals. 
(c) Results at $W=17.5$ (cp.\ Fig.\ \ref{fig-bcc-E-scal}) for $3$ out of $8$ best fit models with $108$ non-equally spaced data points used in each FSS procedure. 
The numerical fitting procedure continued in all cases until convergence was reached or (a) $5000$, (b,c) $1000$ iterations had been completed.}
\flushleft{(a)}\\[2ex]
\begin{tabular*}{\hsize}{@{\extracolsep{\fill}}cccccccccc} \hline\hline
    $\Delta M$  & $E$  & $\Delta W$& $n_{\rm r}$ & $n_{\rm i}$ & $m_{\rm r}$ & $m_{\rm i}$  & $W_{\rm c}$ & $\nu$ & $y$ \\ \hline
    3 - 15 & 0 & 20.3 - 21.5 & 2 & 0 & 1 & 0 & 20.95(1) & 1.67(5) & - \\ 
    3 - 15 & 0 & 20.3 - 21.5 & 3 & 1 & 1 & 4 & 20.92(2) & 1.51(9) & 1.7(5) \\ 
    3 - 15 & 0 & 20.3 - 21.5 & 3 & 2 & 3 & 1 & 20.75(3) & 1.70(9) & 3.0(5) \\
\vdots & \vdots & \vdots & \vdots & \vdots & \vdots & \vdots & \vdots & \vdots & \vdots \\ \hline
\multicolumn{7}{l}{averages:} & 20.85(1) & 1.61(1) & \\ \hline 
\end{tabular*}
\\[3ex]
\flushleft{(b)}\\[2ex]
\begin{tabular*}{\hsize}{@{\extracolsep{\fill}}cccccccccc}
\hline\hline
    $\Delta M$  & $\Delta E$& $W$  & $n_{\rm r}$ & $n_{\rm i}$ & $m_{\rm r}$ & $m_{\rm i}$  & $E_{\rm c}$ & $\nu$ & $y$ \\ \hline    
9 - 13 & 9.9 - 10.9 & 15 & 2 & 0 & 1 & 0 & 10.38(1) & 1.32(5) & -\\
9 - 13 & 9.9 - 10.9 & 15 & 2 & 0 & 2 & 0 & 10.38(1) & 1.22(5) & -\\
9 - 13 & 9.9 - 10.9 & 15 & 3 & 0 & 4 & 0 & 10.40(1) & 1.03(3) & -\\
\vdots & \vdots & \vdots & \vdots & \vdots & \vdots & \vdots & \vdots & \vdots & \vdots \\ \hline
\multicolumn{7}{l}{averages:} & 10.39(1) & 1.27(1) & \\ \hline 
%
\end{tabular*}
\\[3ex]
\flushleft{(c)}\\[2ex]
\begin{tabular*}{\hsize}{@{\extracolsep{\fill}}cccccccccc}
\hline\hline
    $\Delta M$  & $\Delta E$& $W$  & $n_{\rm r}$ & $n_{\rm i}$ & $m_{\rm r}$ & $m_{\rm i}$  & $E_{\rm c}$ & $\nu$ & $y$ \\ \hline    
7 - 15 & 10.5 - 11.5 & 17.5 & 2 & 0 & 1 & 0 & 10.98(1) & 1.55(6) & -\\
7 - 15 & 10.5 - 11.5 & 17.5 & 3 & 0 & 2 & 0 & 10.99(1) & 1.48(6) & -\\
7 - 15 & 10.5 - 11.5 & 17.5 & 3 & 0 & 4 & 0 & 10.99(1) & 1.36(7) & -\\
\vdots & \vdots & \vdots & \vdots & \vdots & \vdots & \vdots & \vdots & \vdots & \vdots \\ \hline
\multicolumn{7}{l}{averages:} & 10.96(3) & 1.47(9) & \\ \hline \hline
\end{tabular*}
\label{tab:bcc}
\end{table}
gives some examples of models providing the best fits and the resulting critical parameters. The values of the critical disorder and critical exponent obtained by averaging over all the best fit models are also given.

\begin{figure}
  \centering
  \includegraphics[width=\figwidth]{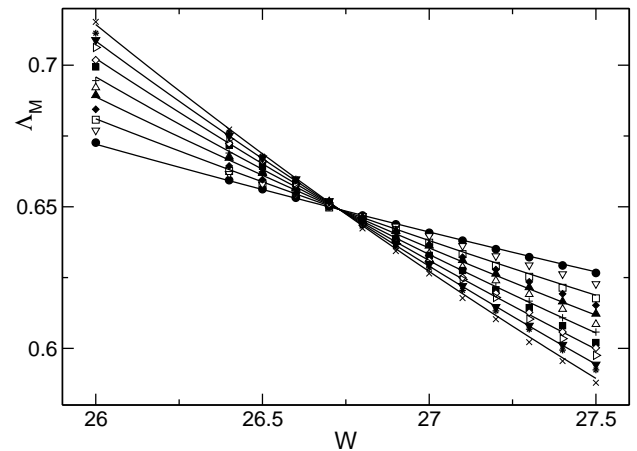}
  \caption{
   Reduced localization lengths $\Lambda_M$ versus disorder $W$ for FCC lattice. 
   Symbol sizes $M$ are $3 (\bullet), 4 (\triangledown), \ldots, 15 (\times)$. Error bars are within symbol size. Lines are fits to the data given by Eqs.\ (\ref{eq-Slevin}) -- (\ref{eq-Slevin-Var})
   with $n_{\rm r}=2, m_{\rm r}=2$. Lines for even $M$ have been removed for clarity. }
\label{fig-fcc-data}
\end{figure}

\begin{figure}
  \centering
  \includegraphics[width=\figwidth]{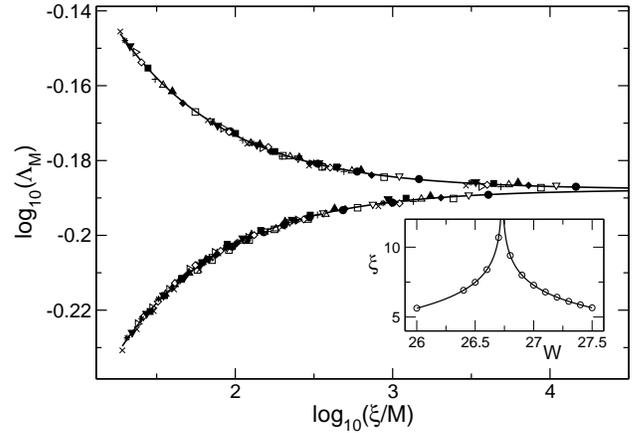}
  \caption{
   Scaling function (solid line) and scaled data points for the FCC lattice and
   $n_{\rm r}=2, m_{\rm r}=2$. Symbols denote the same values of $M$ as in Fig.\ \ref{fig-fcc-data}. Inset: Dependence of the scaling parameter $\xi$ on the
    disorder strength $W$ for the $13$ $W$ values shown in Fig.\ \ref{fig-fcc-data}. In all cases error bars are within symbol size. }
\label{fig-fcc-scal}
\end{figure}

Results of the TMM calculations for the FCC lattice are shown in Fig.\ \ref{fig-fcc-data}. 
In this case the lines for constant $M$ cross at the same point --- at least within the accuracy of the calculated $\Lambda_M$ --- indicating that the use of the irrelevant variables in Eqs.\ (\ref{eq-Slevin2}) and  (\ref{eq-Slevin-Var}) is not necessary in most cases and $n_{\rm i}=m_{\rm i}=0$. Results of the fit for $n_{\rm r}=2, m_{\rm r}=2$ are displayed in Fig.\ \ref{fig-fcc-scal}.
The transition at $W \approx 26.73$ is clearly indicated. More examples of best fit models can be found in Table \ref{tab:fcc} (a) as well as the average values of the critical parameters.
\begin{table}
  \centering
\caption{Critical parameters for the MIT in the FCC lattice. The fitting procedure was continued until convergence was reached or until $1000$ iterations had been completed. 
(a) At $E=0$ (cp.\ Fig.\ \ref{fig-fcc-data}), we use $91$ data points for each FSS and the obtained $31$ best fit models average as indicated. (b) $83$ data points (cp.\ Fig.\ \ref{fig-fcc-E-scal}) are used to perform the FSS at $W=18$ and $8$ best fit models are used to obtain the averages.}
\flushleft{(a)}\\[2ex]
  \begin{tabular*}{\hsize}{@{\extracolsep{\fill}}cccccccccc} \hline\hline
    $\Delta M$  & $E$  & $\Delta W$& $n_{\rm r}$ & $n_{\rm i}$ & $m_{\rm r}$ & $m_{\rm i}$  & $W_{\rm c}$ & $\nu$ & $y$ \\ \hline
    3 - 15 & 0 & 26 - 27.5 & 1 & 4 & 3 & 4 & 26.72(1) & 1.49(4) & 6(2)\\ 
    3 - 15 & 0 & 26 - 27.5 & 2 & 0 & 2 & 0 & 26.73(1) & 1.58(3) & - \\ 
    3 - 15 & 0 & 26 - 27.5 & 3 & 4 & 2 & 4 & 26.72(1) & 1.51(10) & 4(2) \\
\vdots & \vdots & \vdots & \vdots & \vdots & \vdots & \vdots & \vdots & \vdots & \vdots \\ \hline
\multicolumn{7}{l}{averages:} & 26.72(1) & 1.53(1) \\ \hline 
\end{tabular*}
\\[3ex]
\flushleft{(b)}\\[2ex]
\begin{tabular*}{\hsize}{@{\extracolsep{\fill}}cccccccccc}
\hline\hline
$\Delta M$  & $\Delta E$& $W$  & $n_{\rm r}$ & $n_{\rm i}$ & $m_{\rm r}$ & $m_{\rm i}$  & $E_{\rm c}$ & $\nu$ & $y$ \\ \hline
9 - 15 & 8.52 - 8.88 & 18 & 1 & 0 & 2 & 0 & 8.683(3) & 1.63(5) & -\\
9 - 15 & 8.52 - 8.88 & 18 & 2 & 0 & 2 & 0 & 8.687(4) & 1.65(5) & -\\
9 - 15 & 8.52 - 8.88 & 18 & 3 & 0 & 1 & 0 & 8.685(3) & 1.62(5) & -\\
\vdots & \vdots & \vdots & \vdots & \vdots & \vdots & \vdots & \vdots & \vdots & \vdots \\ \hline
\multicolumn{7}{l}{averages:} & 8.684(2) & 1.63(2) & \\ \hline \hline
\end{tabular*}
\label{tab:fcc}
\end{table}

\subsection{Critical Parameters away from the Band Centre}
\label{sec-criteng}

\begin{figure}
  \centering
  \includegraphics[width=\figwidth]{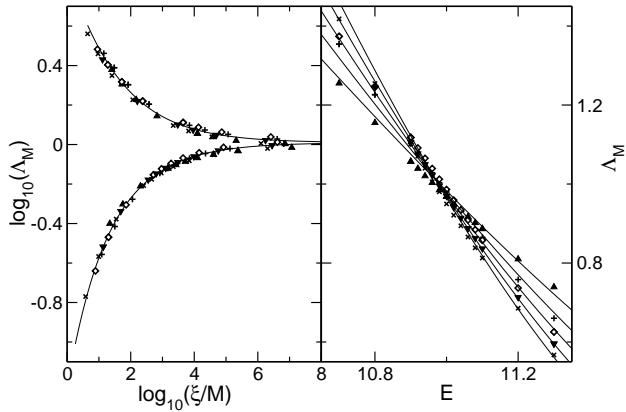}
  \caption{
Localization data for the BCC lattice with $W=17.5$. System sizes $M$ are $7 (\blacktriangle), 9 (+), \ldots, 15 (\times)$ as in Fig.\ \ref{fig-bcc-data}. Error bars are within symbol size. Left: Scaling function (solid line) and scaled data points using
   $n_{\rm r}=2, m_{\rm r}=1$. Right: Reduced localization lengths $\Lambda_M$ versus disorder $W$. Lines are fits to the data given by Eqs.\ (\ref{eq-Slevin}) -- (\ref{eq-Slevin-Var})
    with $n_{\rm r}=2, m_{\rm r}=1$.  }
\label{fig-bcc-E-scal}
\end{figure}

We also perform calculations where we fix the disorder and allow the energy to vary across a critical value $E_{\rm c}$ for the transition. We remark that it is know that such investigations are numerically more difficult due to the influence of density of states effects.\cite{CaiRS99} Results of the TMM and FSS calculations for the BCC lattice with $W = 17.5$ can be seen in Fig.\ \ref{fig-bcc-E-scal}. 
Although the lines for constant $M$ do not all cross at the same point, the data is best fitted by models containing no irrelevant scaling. Also evident is the poorer quality of the fit compared to the calculations where the energy was fixed at zero. This is not due to using lower accuracy data, as the maximum raw-data error remained at $0.1 \%$. Hence we attribute it to complications arising from a varying density of states close to $E_{\rm c}$ at the attainable values of $M$. Results for the critical parameters are shown in Table \ref{tab:bcc} (b) and (c) for $W=15$ and $W=17.5$ respectively. The low value of $\nu$ for the case $W=15$ can be attributed to the use of fewer data points in the FSS and only using three values of $M$. We note that this is consistent with the lower values of $\nu$ obtained in the diagonalization studies as mentioned in the Introduction. It appears that the FSS procedure systematically reduces the values of the critical exponent for data from smaller systems or of lower accuracy.

\begin{figure}
  \centering
  \includegraphics[width=\figwidth]{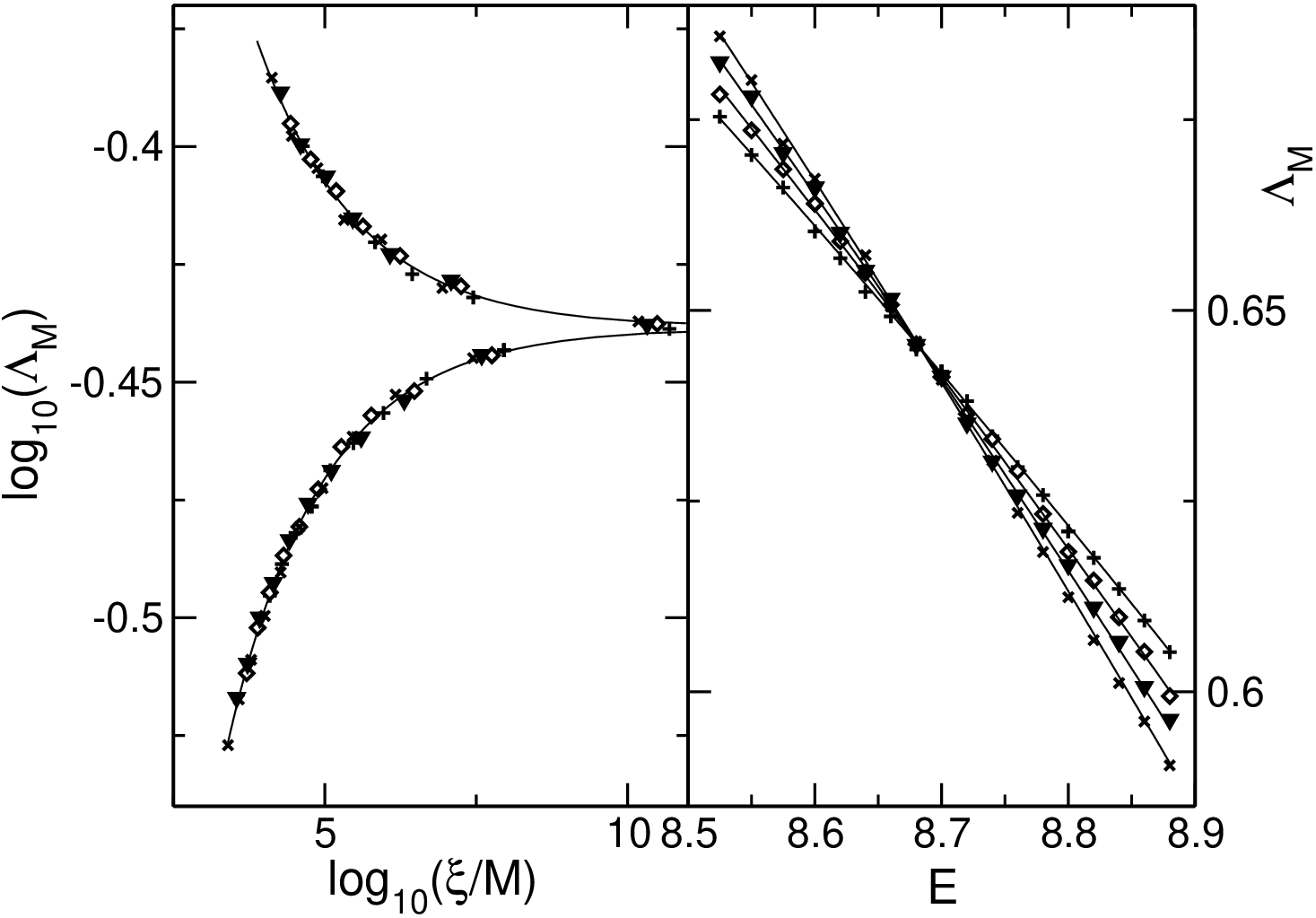}
  \caption{
Data for the FCC lattice with $W=18$. System sizes $M$ are $9 (+), 11 (\Diamond), \ldots, 15 (\times)$ as in Fig.\ \ref{fig-fcc-scal}. Error bars are within symbol size. Left: Scaling function (solid line) and scaled data points using
   $n_{\rm r}=1, m_{\rm r}=2$. Right: Reduced localization lengths $\Lambda_M$ versus disorder $W$. Lines are fits to the data given by Eqs.\ (\ref{eq-Slevin}) -- (\ref{eq-Slevin-Var})
    with $n_{\rm r}=1, m_{\rm r}=2$.  }
\label{fig-fcc-E-scal}
\end{figure}

Results for the TMM and FSS calculations for the FCC lattice with $W=18$ can be seen in Fig.\ \ref{fig-fcc-E-scal}.
Table \ref{tab:fcc} (b) gives examples of the best fit models and shows the resulting average critical parameters. Note that both estimates of $\nu$ are consistent with the result $1.57(2)$ for the SC lattice.\cite{SleO99a}

\section{Conclusions}
\label{sec-concl}
Using the transfer-matrix approach and FSS we determined the critical parameters of the Anderson transition for the BCC and FCC lattices. The values of the critical exponent $\nu$ are in good agreement with the results obtained previously for other systems belonging to the orthogonal universality class. The increase of the critical disorder $W_{\rm c}$ from $16.54$ for the SC lattice to $20.85$ for BCC and $26.72$ for FCC lattice may be attributed to an increasing number of nearest neighbors which for the above structures equals $6$, $8$ and $12$, respectively. More nearest neighbors connected to a given site provide more paths for electronic transport, so stronger disorder is needed to localize eigenstates of the system. The universal localization properties of a 3D system and the presence of an MIT are however not affected in accordance to the scaling theory of localization \cite{AbrALR79} and in agreement with results \cite{SchO92} showing that they depend only on the dimensionality of the system, but not on the number of nearest neighbors in the lattice. 

Our results and their interpretation are consistent with investigations of classical bond and site percolation models on SC, BCC and FCC lattices. In Ref.\ \onlinecite{GalM96} it was found that the percolation thresholds for these lattices decrease with increasing number of nearest neighbors; more neighbors allow for easier formation of a percolating cluster, or, as in our case, the formation of extended states.

\acknowledgments
We thankfully acknowledge discussions with C.\ Hooley, J.\ Knoester and I.\ Plyushchay. We are grateful to T.\ Wright for producing the $M=7$ and $9$ data for the phase diagrams. We thank S.\ Wells for careful reading of the manuscript. AMF and RAR gratefully acknowledge EPSRC for financial support.

\appendix

\section{Connectivity Matrices}
\label{sec-connectivity}

For completeness, let us give the connectivity matrices for BCC and FCC lattices with $M=3$. Recall that $\bm{\mathrm{C}}_{l}$ is the connectivity matrix describing the connections of the $l$th slice to the $l-1$th slice. Element $c_{jk}$ of the connectivity matrix  equals $1$ if site $j$ in the $l$th slice is connected to site $k$ in the $l-1$th slice; otherwise $c_{jk}=0$. The boundary terms are indicated in italics.
For the BCC lattice for odd layers,
\begin{equation}
\label{eq-cibccodd}
\boldsymbol{\mathrm{C}}_{2l-1} = 
\left( \begin{array}{ccccccccc}
1 & 0 & {\it 1} & 0 & 0 & 0 & {\it 1} & 0 & {\it 1} \\
1 & 1 & 0 & 0 & 0 & 0 & {\it 1} & {\it 1} & 0 \\
0 & 1 & 1 & 0 & 0 & 0 & 0 & {\it 1} & {\it 1} \\
1 & 0 & {\it 1} & 1 & 0 & {\it 1} & 0 & 0 & 0 \\
1 & 1 & 0 & 1 & 1 & 0 & 0 & 0 & 0 \\
0 & 1 & 1 & 0 & 1 & 1 & 0 & 0 & 0 \\
0 & 0 & 0 & 1 & 0 & {\it 1} & 1 & 0 & {\it 1} \\
0 & 0 & 0 & 1 & 1 & 0 & 1 & 1 & 0 \\
0 & 0 & 0 & 0 & 1 & 1 & 0 & 1 & 1 \end{array} \right).
\end{equation}
For even layers
\begin{equation}
\label{eq-cibcceve}
\boldsymbol{\mathrm{C}}_{2l} = 
\left( \begin{array}{ccccccccc}
1 & 1 & 0 & 1 & 1 & 0 & 0 & 0 & 0 \\
0 & 1 & 1 & 0 & 1 & 1 & 0 & 0 & 0 \\
{\it 1} & 0 & 1 & {\it 1} & 0 & 1 & 0 & 0 & 0 \\
0 & 0 & 0 & 1 & 1 & 0 & 1 & 1 & 0 \\
0 & 0 & 0 & 0 & 1 & 1 & 0 & 1 & 1 \\
0 & 0 & 0 & {\it 1} & 0 & 1 & {\it 1} & 0 & 1 \\
{\it 1} & {\it 1} & 0 & 0 & 0 & 0 & 1 & 1 & 0 \\
0 & {\it 1} & {\it 1} & 0 & 0 & 0 & 0 & 1 & 1 \\
{\it 1} & 0 & {\it 1} & 0 & 0 & 0 & {\it 1} & 0 & 1 \end{array} \right).
\end{equation}
For the FCC lattice for odd and even layers
\begin{equation}
\label{eq-cifcc}
\boldsymbol{\mathrm{C}}_{l} = 
\left( \begin{array}{ccccccccc}
1 & 1 & 0 & 0 & 0 & 0 & 0 & {\it 1} & 0 \\
0 & 1 & 1 & 0 & 0 & 0 & 0 & 0 & {\it 1} \\
{\it 1} & 0 & 1 & {\it 1} & 0 & 0 & 0 & 0 & 0 \\
0 & 1 & 0 & 1 & 1 & 0 & 0 & 0 & 0 \\
0 & 0 & 1 & 0 & 1 & 1 & 0 & 0 & 0 \\
0 & 0 & 0 & {\it 1} & 0 & 1 & {\it 1} & 0 & 0 \\
0 & 0 & 0 & 0 & 1 & 0 & 1 & 1 & 0 \\
0 & 0 & 0 & 0 & 0 & 1 & 0 & 1 & 1 \\
{\it 1} & 0 & 0 & 0 & 0 & 0 & {\it 1} & 0 & 1 \end{array} \right).
\end{equation}
In all cases, $l=1, 2, \ldots$\ .

%
%


\begin{thebibliography}{10}
\bibitem{minussign} Note the minus in the top-left entry of the matrix in Eq.\ (\ref{eq-tmmred}); this is due to the minus before the hopping term in Eq.~(\ref{hamilt}).
\bibitem{barycenter} Let $\bm{\mathrm{H}}$ be similar to some diagonal matrix $\bm{\mathrm{H}_0}$, so that $\bm{\mathrm{H}}=\bm{\mathrm{U}}\bm{\mathrm{H}_0}\bm{\mathrm{U}^{-1}}$, where $\bm{\mathrm{U}}$ is a unitary matrix. Let $\{E_i\}$ denote the eigenvalues of $\bm{\mathrm{H}_0}$, which are equal to the eigenvalues of $\bm{\mathrm{H}}$. Then 
$\sum_{i} E_i={\rm Tr}(\bm{\mathrm{H}_0})={\rm Tr}(\bm{\mathrm{H}})=\sum_{i}\epsilon_{i}\approx0$ and the average eigenvalue of $\bm{\mathrm{H}}$ is approximately equal to $0$.
\end{thebibliography}

\end{document}